\documentclass[preprint,prx,longbibliography]{revtex4-1}
\usepackage{subcaption}
\usepackage{graphicx}
\usepackage{comment}
\usepackage{amsmath,amssymb,amsthm} 
\usepackage{scalerel,stackengine}
\usepackage{bm}
\usepackage{verbatim}
\usepackage{float}
\usepackage{color,soul}
\usepackage{textcomp}
\usepackage{rotating}
\usepackage{color,soul}
\usepackage{mathrsfs}
\usepackage[title]{appendix}
\usepackage{xcolor}

\begin{document}
\title{Approximations and modifications of celestial dynamics tested on the three-body system  }
\author{ S\o ren  Toxvaerd }
\affiliation{ Department of Science and Environment, Roskilde University, Postbox 260, DK-4000 Roskilde, Denmark}
\email{To appear in Celest. Mech. Dyn. Astron. 2026, Email: st@ruc.dk}

\begin{abstract}
Celestial Newtonian systems have regular dynamics, but with
classical Keplerian rotation velocities, which disagree with the observed rotation of galaxies in the Universe. However,
modifications of the classical accelerations or gravitational attractions can overcome this defect. 
The large-scale simulations of galaxies are with approximations with ``particle-mesh'' (PM) substitutions of
the attractions from objects far away, which affects the regular dynamics. Here, we investigate the impact of the PM approximation and of the modifications
of accelerations or gravitational attractions on the stability of the regular dynamics in a celestial system.
The simple three-body system (TBS) is the simplest system to test the stability of the regular dynamics with
approximations or with modifications of celestial dynamics,
and it is easy to implement on a computer. Simulations of the TBS show that
the PM approximation, and the modification of the accelerations (MOND), destabilizes TBS.
In contrast, a modification of gravity by replacing  Newton's inverse square attraction
with an increased attraction (Yukawa, MOGA) for far-away interactions stabilizes the system.
The PM approximation and the MOND modification of classical dynamics do not preserve
the momentum and angular momentum of a conservative system exactly, and they do not obey Newton's third law. Although these errors
and shortcomings are small, they  eventually cause the instability of the regular dynamics.
\end{abstract}

\maketitle
	    
\section{Introduction}
The time evolution of astronomical systems is obtained by simulations, where the objects' positions are 
 calculated employing algorithms for discrete classical dynamics. This approach has been used since Newton's time,
 and today, the dynamics of galaxies with billions of suns are obtained by large-scale simulations. An overview of simulations of
 galaxial systems is given in  \cite{Springel2016,Vogelsberger2020}.
The dynamic behavior of these large-scale simulations is achieved by making  ``particle-mesh'' approximations (PM),
PM3 \cite{Hockney1974,Epstathiou1985} or Tree-PM \cite{Bagla2002}, for the small effects of
the many objects that are far away and which are assumed to have only a minor influence on an object's motion.
However, recent simulations indicate that these approximations used in the simulations can destabilize the systems \cite{Toxvaerd2025a}.

Large-scale simulations of galaxy models show that the simulated galaxies are unstable and have a classical Kepler-like rotation
that deviates from the experimentally determined rotation velocities of galaxies. This inspired Milgrom to formulate a classical
dynamics with a modified acceleration at large interstellar distances (MOND) \cite{Milgrom1983},
 and the MOND modification has later been reformulated (QuMOND) \cite{Milgrom2010}.
Another way to overcome the shortcomings in the simulated galaxies is to modify the gravitational
attractions at large interstelar distances   \cite{Fischbach2001,Adelberger2003,Brandau2012,Toxvaerd2024a}.

Here, we investigate the impact of the particle-mesh approximations and
the modifications of Newtonian dynamics on the stability of the regular dynamics in a three-body system (TBS). The investigation is
 for a  TBS of a simple Dwarf-galaxy, consisting of a heavy mass center (``Black hole''), and two objects in regular orbits around the center. The regular dynamics without
 approximations or modifications are shown in Figure 1, and the mean
distances  and units are given in Table 1. 

\begin{figure}
 \begin{center}	  
 	 \includegraphics[width=6cm,angle=-90]{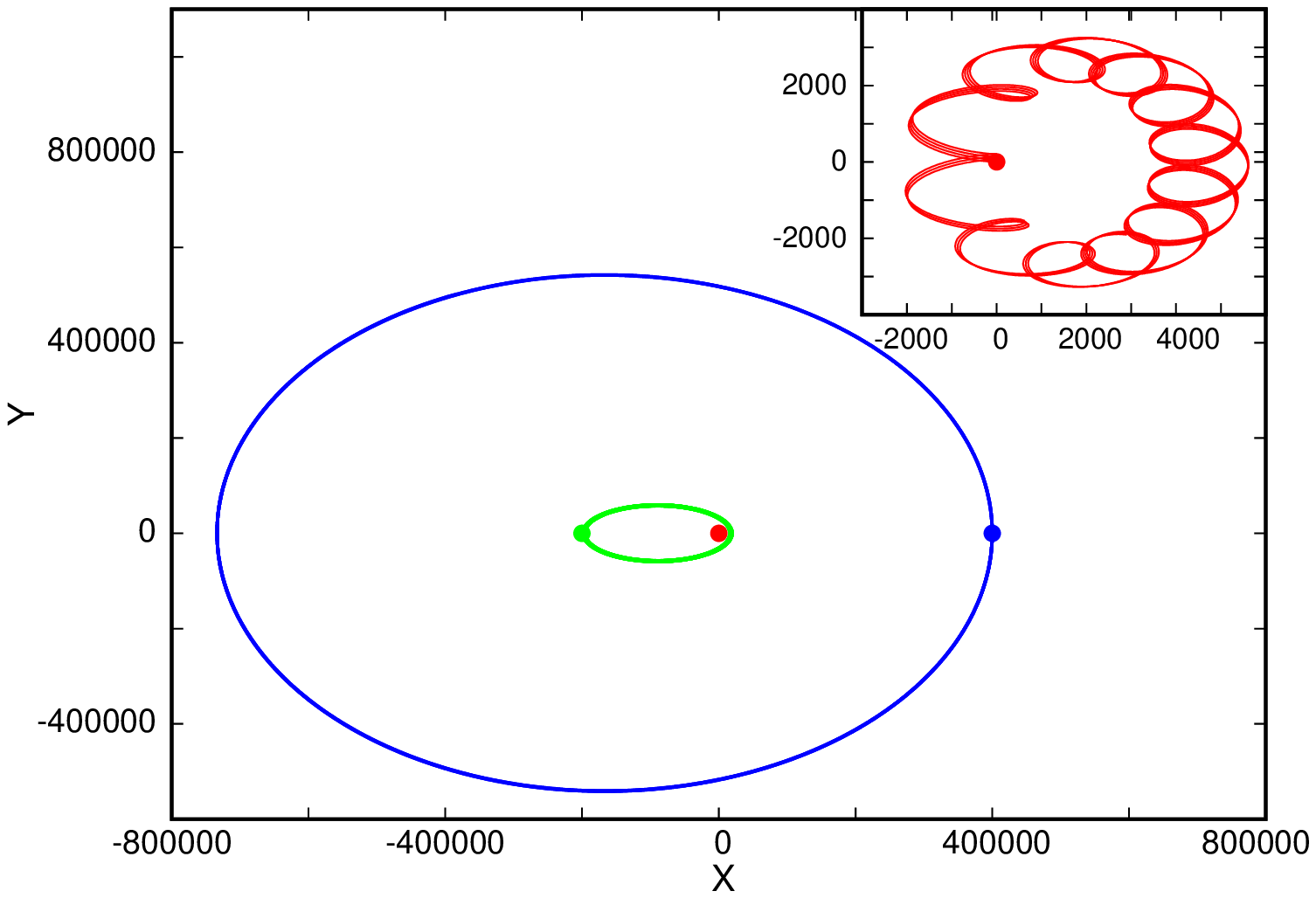}
	  \caption{The regular orbits in a  classical mechanics TBS system with two light objects around a
	  heavy object, and for $10^7$ discrete timesteps. The $\approx$ 45 orbits in green is
	 for the object No. 1 with start position at its aphelion (green dot), and with blue is the corresponding  $\approx$  four orbits
	 for the object No. 2 with start position at its perihelion (blue dot). The red ``dot'' is the orbits
	 of the heavy object No. 3  with start position
	 at the center of mass (origin), and the inset shows the  $\approx$ four orbits of the object.}
 \end{center}	
\end{figure}

The TBS has challenged science ever since Newton formulated classical mechanics in 1687,
and simultaneously solved the dynamics of two celestial bodies \cite{Newton1687}.
Newton's solution of the dynamics of the two-body system
satisfies Kepler's three laws for the planets in the solar system. Since the planets in the Solar system move in orbits, bound rotations must be
a classical mechanical solution for a system with three as well as many celestial bodies. But although one cannot
solve the coupled second-order differential equations, even for just a TBS, regular solutions have been
found for the system \cite{Suvakov2013,Dmitra2028,Li2025}. An overview of
the history of the three-body system and its dynamic behavior is given in \cite{Krishnaswami2019},  and how to simulate a TBS system
are given in the Appendix.

\begin{table}
\caption{Data for the three-body  ``Dwarf galaxy'' model.}
\begin{tabbing}
\hspace{4.6cm}\=\hspace{2.5cm}\=\hspace{2.6cm}\=\hspace{2.6cm}\=\hspace{2.6cm} \\
Object \>   Mean distance   \>  Mass  \>  Mean distance   \>  Mass \\

	No. 1 (``Sun'')  \> 145000  \>  1  \>  83400\footnotemark[2]   \>   $M_{\odot}$=1  \\
	No. 2 (Outer star)   \> 590000    \>  0.5    \>     \>   \\
	No. 3 ( ``Black hole'')   \>  3500  \>  100   \>  0  \>   \\
\end{tabbing}
	\footnotetext{ Units in the TBS model: Mass  in unit of the No. 1's (``Sun'')  mass   $M_{\odot}$=1, length unit in pc  \cite{Toxvaerd2022a}, and the gravitational constant $G=$1. }
	\footnotetext[1]{For set up of units in the model see \cite{Toxvaerd2022a}.}
	\footnotetext[2]{Distances in the Milky Way \cite{Reid2014}.}
\end{table}
Simulations of systems with classical dynamics are performed using discrete algorithms,
and almost all simulations are with Newton's discrete algorithm \cite{Newton1687,Toxvaerd2023}.
The classical dynamics obtained with this algorithm is time-reversible and symplectic, and has the same invariances (momentum, angular momentum, and energy)  as the analytic dynamics. Hence, it is exact in the same sense as the corresponding exact
solution to the coupled second-order differential equations for Newton's analytic dynamics \cite{Toxvaerd2023,Toxvaerd2024,Toxvaerd2025}.
Here, the exact discrete dynamics for TBS are used to
investigate the sensitivities of the regular orbits to the various approximations and modifications
used in large-scale simulations of celestial systems' dynamics.

\section{The dynamics of a three-body system}\label{sec2}

 Newtonian dynamics for celestial objects is given by Newton's classical dynamics and his inverse-square law (ISL) of  gravitation  
\begin{equation}
	\textbf{F}_{i}(t)=m_i \textrm{a}_i \hat{\textbf{a}}_i(t)= \sum_{j \ne i}^N \textbf{f}_{ij}(t)= 
	-\sum_{j \ne i}^N {\frac{G m_i m_j}{r_{ij}^2(t)} \hat{\textbf{r}}_{ij}(t)}
\end{equation}	
for  the  acceleration $ \textbf{a}_i(t)= \textrm{a}_i \hat{\textbf{a}}_i(t)$  caused by the sum of forces $\sum_{j \ne i}^N  \textbf{f}_{ij}(t)$
 on the object $i$ in the ensemble of $N$ objects,
 by  baryonic objects $j$ with mass $m_j$ at  distances $r_{ij}(t)$,  and at time $t$.

   \begin{figure}
\begin{center}   
	   
 	 \includegraphics[width=6cm,angle=-90]{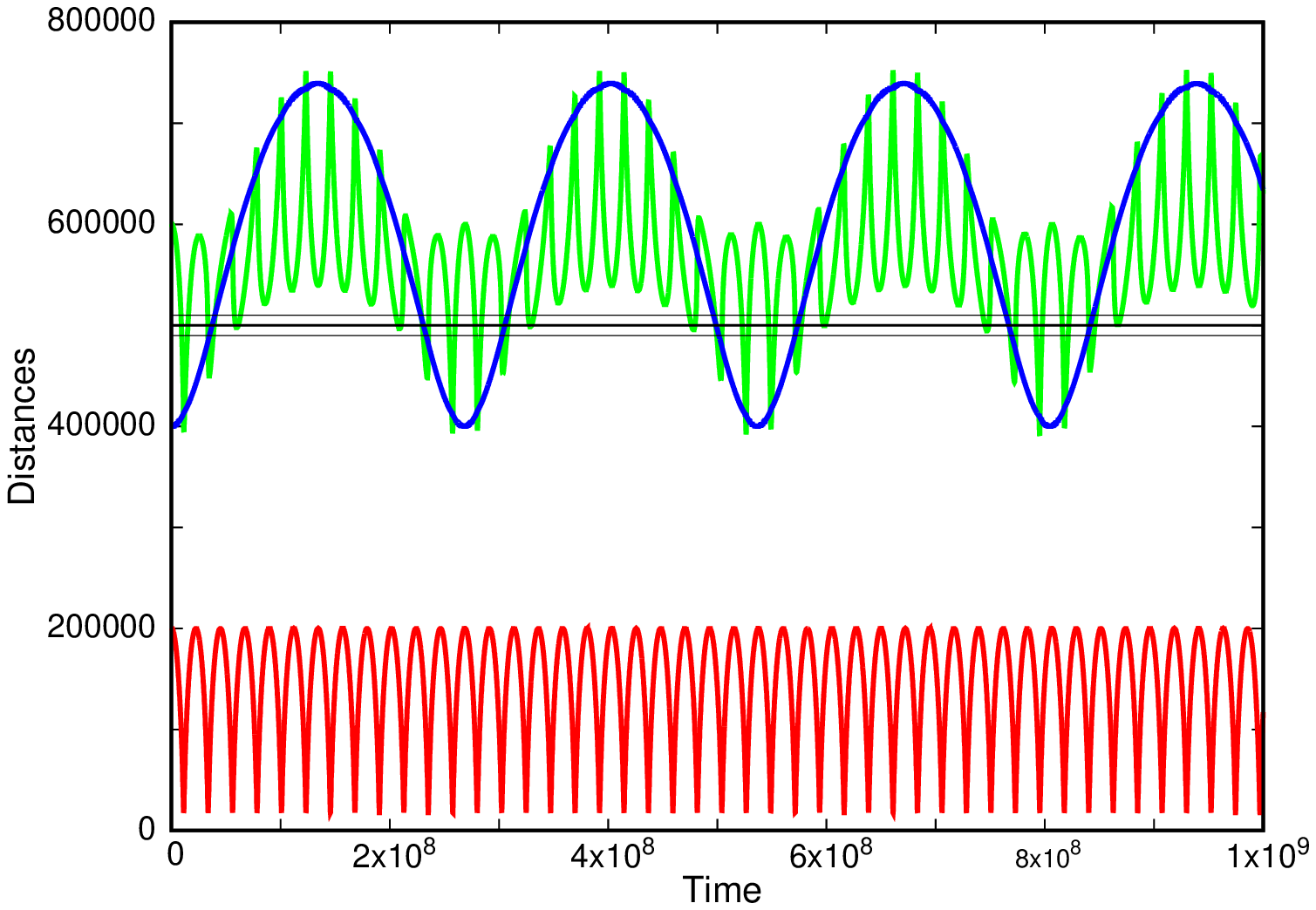}
	   \caption{ The distances $r_{12}(t)$ (green),  $r_{13}(t)$ (red), and  $r_{23}(t)$ (blue) between the three objects as a function of time
	   for the regular orbits, shown in Figure 1.
 The thick black straight line is the borderline for the PM approximation, $r_0$=500000, and the
	   thinn black lines are  $r_0 \pm l_{grid}$ with $l_{grid}$= 10000, used in the  PM approximation. }   
\end{center}	   
 \end{figure}

 Newton's discrete dynamics for TBS and how to start the TBS are described in the Appendix.
The TBS can have solutions with
  regular orbits of the objects around their center of gravity. Here we want to set up a TBS with a heavy ``sun'' or mass center in a galaxy with
  two planets or stars in a galaxy, one inner and one outer object in the TBS.
  The TBS system with the ellipses in Figure 1 is
 with  
 (mass in units of  objects' No. 1 mass, and dynamics units given by the gravitational constant $G=1$): 
 $m_1=1, m_2$=0.5 and $m_3$=100, and with the start position of  No. 1 at  $[x_1(0),y_1(0)]= [-200000, 0]$, 
 No. 2 at  $[x_2(0),y_2(0)]= [400000,0]$, and
  with the heavy object at the origin at the start $[x_3(0),y_3(0)]=[0,0]$. The orbits in the figure are obtained for $10^7$ time steps with
  $\delta  t$=100 where object No. 1 in green
had circulated $\approx$ 45 times in the elliptical orbit. In blue is the corresponding  $\approx$ four orbits of the light object No. 2, and the inset
shows the heavy object's more complicated four orbits in red. The TBS was simulated $10^{10}$ time steps, and the regular dynamics is stable.

The TBS is used for testing the stability of the system when exposed to the approximations and modifications  of celestial dynamics.

\section{PM approximation of far-away interactions in the  three-body system}
 In  large-scale  simulations of galaxies, one
 sorts the  N-1 forces on object $i$  in Eq. (1)  
  into a sum of   $N_i'(t)$  short-range forces $\textbf{f}_{ij}(r_{ij})$
 with  distances $r_{ij}(t)$, which are less than a certain (large) distance
   $r_0 $, and for which the forces  are computed  directly, and  $N_i''(t)$  forces
 from   objects far away with $r_{ij}(t)>r_0$. Their positions  are interpolated onto a mesh with a certain
 grid length $l_{grid}$, and their forces are taken   from the centers  $\textbf{r}_{\alpha(j)}(t)$
of the boxes $\alpha(j),$ with   grid length $l_{grid}$ and with a distance $r_{i\alpha(j)}(t)$ and  direction
$ \hat{\textbf{r}}_{i\alpha(j)}(t)$ to object $i$. The force on object $j$ from $i$ is corespondingly
taken from the center $\textbf{r}_{\alpha(i)}(t)$ of the subbox with object $i$.

  \begin{figure}
 \begin{center}	  
\begin{subfigure}[a]{1.00\linewidth}
\begin{center}	 
 	 \includegraphics[width=5.9cm,angle=-90]{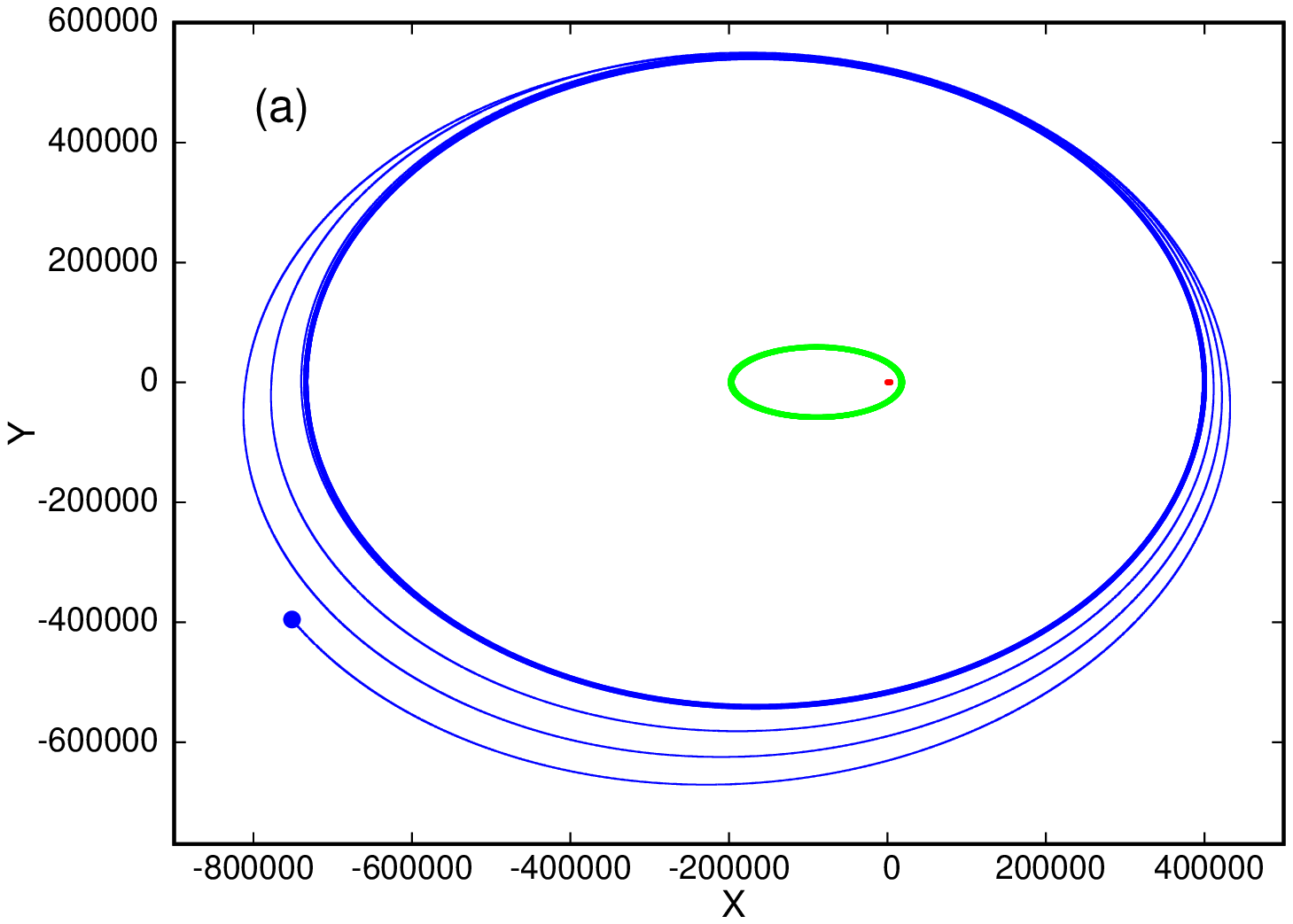}
	 \caption{ $10^7$ timesteps with the PM approximation.}
 \end{center}	
\end{subfigure}

\begin{subfigure}[b]{1.00\linewidth}	  	  
\begin{center}	
 	 \includegraphics[width=5.9cm,angle=-90]{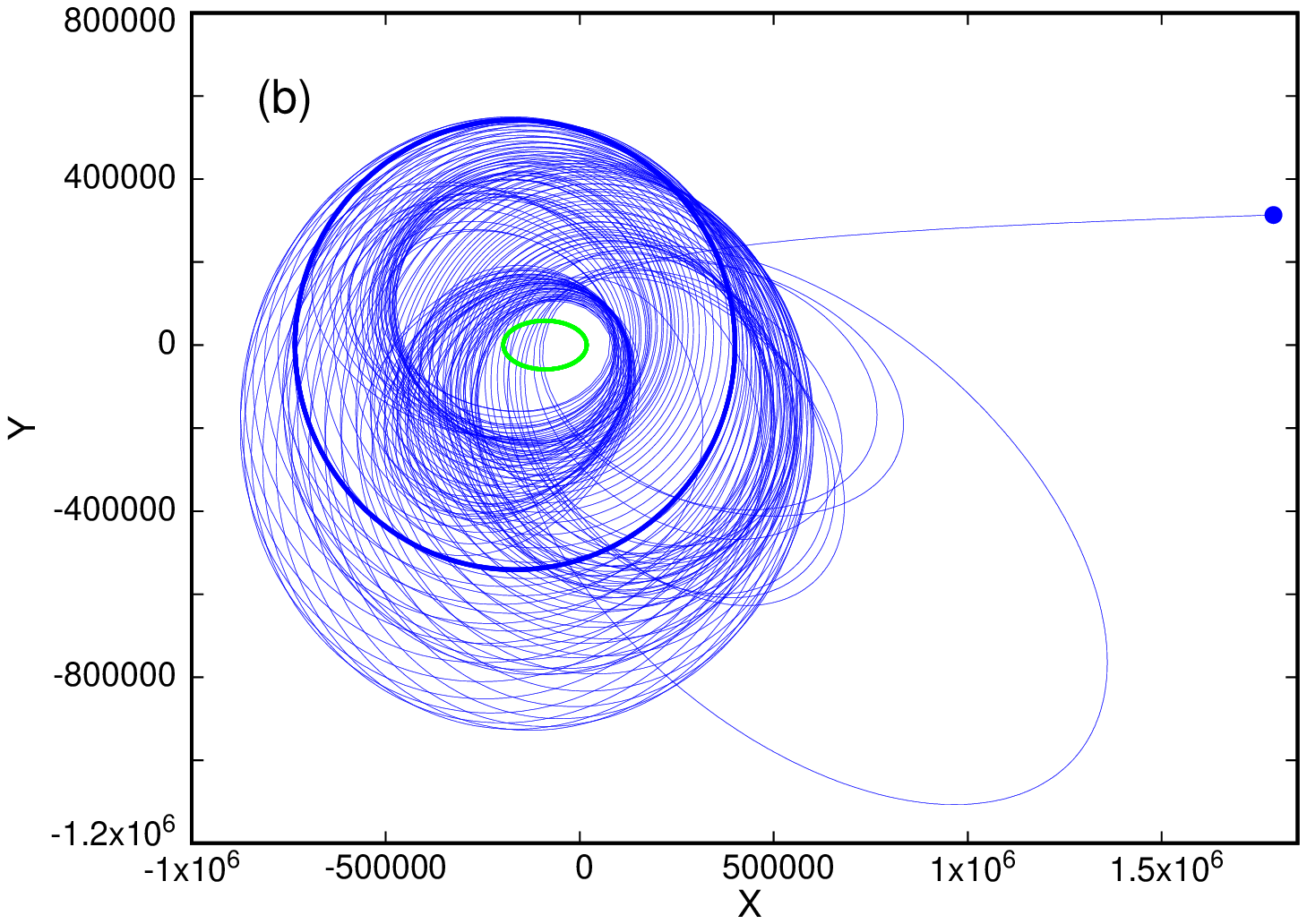}
	 \caption{ 1.49899436$\times 10^8$ timesteps with PM, and when No. 2 was released from the TBS.}
\end{center}	
\end{subfigure}	 
	  \end{center}
	  \caption{The dynamics of TBS with the PM approximation with $r_0=500000$ and $l_{grid}$=10000. The blue orbits are for  No. 2. The 
	  thick blue ellipse is without PM approximation (also shown in Figure 1), and the orbits with thin blue are with the PM approximation.
	  No. 1's orbits are  in green  with or without PM (the differences are not visible on the figures).}
\end{figure}

The division of the space into a sum of cubes with mean field attractions to $i$ has the consequence that 
\begin{eqnarray}
	r_{i\alpha(j)}(t) \neq  r_{\alpha(i)j}(t) \neq  r_{ij}(t)   \\
	\hat{\textbf{r}}_{i\alpha(j)}(t) \cdot \hat{\textbf{r}}_{j\alpha(i)}(t)  \neq -1,
\end{eqnarray}	
and
\begin{equation}
	\textbf{f}_{i\alpha(j)}(t) \neq -\textbf{f}_{j\alpha(i)}(t) \neq \textbf{f}_{ij}(t).
\end{equation}

Eqn.(2)-(3) with $0 \ll  r_0$ and with the size length $l_{grid} \ll r_0 $ looks like  excellent approximations.
But it has the consequence that the symmetry of pair interactions is broken and that Newton's third law
\begin{equation}
	\textbf{f}_{ij}(t) = -\textbf{f}_{ji}(t)
\end{equation}
 is no longer strictly valid
for dynamics with the  PM approximation of the long-range forces.
  The
 momentum and angular momentum of the system are not exactly conserved, and PM might destabilise the regular dynamics in a celestial system \cite{Toxvaerd2025a}.

\subsection{Testing PM approximationen on the TBS}

The TBS system can be used to illustrate the destabilization of a celestial system by the PM approximation, and PM is easily implemented in TBS (see 
 Appendix). PM primarily affects objects at distances in the outer edge of a 
celestial system and in the TBS system object no. 2.
 Figure 2 shows the distances $r_{12}(t), r_{13}(t),$ and $r_{23}(t)$ for the regular orbits without PM, shown in Figure 1.
 With green and bue is $ r_{12}(t), r_{13}(t)$, respectively, and  $ r_{23}(t)$ is in red.  
 The effect of a PM approximation on the regular dynamics of the TBS system, shown in Figure 1, is obtained for a
 far-away distance  $r_0=500000$ between two objects. The far-away distance is $r_0=500000$, marked by a solid line in Figure 2, and
 the thin solid lines are $r_0=500000 \pm l_{grid}$ with $l_{grid}=10000$. This PM choice affects the far-away object, No. 2's
 regular dynamics by changing the direction primarily, and a percent change in the distance on average. However, it is sufficient to disrupt the system's stability.

 Figure 3a and Figure 3b illustrate the effect of the PM approximation with  $r_0=500000$ and  $l_{grid}=10000$. 
  Figure 3a shows the PM effect for the first $10^7$ time steps (the color of the orbits is the same as in Figure 1). 
  The PM affects primarily object No. 2's regular dynamics (in blue),
   and the effect on object No. 1 (in green) and the heavy object No. 3 (in red) is not visible in the figures.
  Object No. 2 begins to perform ``revolving orbits'' by being exposed to the PM, and 
  it is released from its regular dynamics after  1.49899436$\times 10^8$ time steps (Figure 3b).

 The PM effect was tested for many other choices of $r_0$ and $l_{grid}$, and for other TBS systems. The destabilizing decreases with increasing $r_0$ and decreasing
 grid length  $l_{grid}$. But even a small change of focus for the distance $r_{12}$ and  $r_{13}$, and direction of the forces $\textbf{f}_{12}(t)$ and  $\textbf{f}_{23}(t)$ 
 are sufficient to destroy object No's regular dynamics, as illustrated in Figure 3b.

\section{Modifications of accelerations and gravitational attractions}
Theoretically, one can 
modify the classical dynamics of a celestial system of $N$ objects, with the acceleration $\textbf{a}_i$ and force $\textbf{F}_i$ 
\begin{equation}
	\textbf{a}_i(t)=\textbf{F}_i(t)/m_i
\end{equation} in Eq. (1)
 in at least two ways:
by modifying the accelerations $\textbf{a}_i(t)$: MOND,QuMOND \cite{Milgrom1983,Milgrom2010} or by modifying the gravitational
attractions $\textbf{F}_i(t)$: Yukawa-modification  \cite{Fischbach2001}, and MOGA \cite{Toxvaerd2024a}.
 The two kinds of modifications  are qualitatively equal for a system of only two objects,
but they differ significantly for a many-body system.

The modification, $\mu $ in MOND,  of the acceleration  $\mid\textbf{a}_i\mid$
is obtained  with the ''standard interpolation function" 
\begin{equation}
	\mu(\textrm{a}/a_0)=\sqrt{ \frac{1}{1+(\frac{a_0}{\textrm{a}})^2}},
\end{equation}
or the interpolation function  proposed in \cite{Gentile2011}
\begin{equation}
	\mu(\textrm{a}/a_0)=\frac{\mid\textrm{a}\mid}{ \mid\textrm{a}\mid+a_0},
\end{equation}
with the MOND modification given by
\begin{equation}
	\textbf{F}_i/m_i	=
	\textbf{a}_i \frac{\mid\textrm{a}_i\mid}{ \mid\textrm{a}_i \mid+a_0},
\end{equation}
and acceleration 
\begin{equation}
	\textbf{a}_i(\textrm{MOND})  =\frac{ \textbf{F}_i}{2m_i}(1+\sqrt{1+4 m_i a_0/\mid\textbf{F}_i\mid}).
\end{equation}	
 MOND does not conserve momentum and angular momentum of a gravitational system because 
 \begin{equation} 
	 \textbf{f}_{ij}(t)\sqrt{1+4 m_i a_0/\mid F_i\mid} \neq   -\textbf{f}_{ji}(t)\sqrt{1+4 m_j a_0/\mid F_j\mid}.
 \end{equation}	
The modifications, Eq. (7) or Eq. (8), change the acceleration
  from the Newtonian  acceleration for $\mid\textbf{a}_i\mid  >> a_0 $ at short distances to  a modified acceleration
\begin{equation}
	\mid\textbf{a}_i(\textrm{MOND})\mid= \sqrt{ \mid\textbf{F}_i\mid a_0/m_i}
\end{equation}
for $ \mid\textbf{a}_i \mid << a_0$.

The asymptotic modified acceleration for an isolated object $i$  
 with only one gravitational interaction, $\mid\textbf{F}_i(r_{ij})\mid =-m_im_jG/r_{ij}^{2}$, with another object, No. $j$,
 is obtained from Eq. (1) and Eq.(10) as
\begin{equation}
	\mid \textbf{a}_i(\textrm{MOND}) \mid= - \frac{ \sqrt{m_jG a_0}}{r_{ij}}.
\end{equation}

MOND is a modification of  Newtonian acceleration. But in the simple case with only two objects,
the modification might as well  be formulated as a modification
of Newton's ISL law of universal gravitational attraction, where the inverse square attraction asymptotically is replaced
with an inverse attraction  (IA), Eq. (13).
If this $mo$dified $g$ravitational $a$ttraction (MOGA) is a  universal law for a many-body system, the gravitational force is modified to 
 
\begin{equation}
	\textbf{F}_i(\textrm{MOGA})=-\sum^N_{j \ne i}\frac{m_i m_j G}{r_{ij}^2}(1+    \frac{r_{ij}}{r_0}) \hat{\textbf{r}}_{ij}(t),
	\end{equation}
and the connection between MOND and MOGA is
\begin{equation}
	r_0=\sqrt{MG/a_0},
\end{equation}
where $M$ is the  mass which changes the acceleration.

Newtonian dynamics and Newton's discrete algorithm (Appendix) is
symplectic and with the dynamical invariances:  momentum, angular momentum, and energy for a conservative system \cite{Toxvaerd2023}. 
Modifications of the gravitational forces and   with Newton's discrete algorithm, maintains these qualities,
whereas MOND does not conserve momentum and angular momentum according to Eq. 11 \cite{Felten1984},\cite{Toxvaerd2024a}. 

 The MOGA modification of Newton's ISL to an inverse attraction for large interaction distance $R$
is an example of the $f(R)$ modification of gravity, introduced in the standard model \cite{Capozziello2011,Clifton2012,Capozziello2024}.
Simulation of large-scale celestial systems with modified gravity is described in \cite{Winther2015} and cosmological tests
of modified gravity are reviewed in \cite{Koyama2016}.

 Modifications of the Newtonian ISL attraction have been proposed for a long time 
 in an attempt to obtain the stability of galaxies
  by a modification of the standard model. Another modification of the ISL gravity is the ``Yukawa modification'' 
 where the Newtonian gravitational ISL potential  is modified by a Yukawa potential  \cite{Fischbach2001,Adelberger2003,Henrichs2021}

 \begin{equation}
	 v_{ij}(r)=u(r_{ij})[1+\alpha exp(-r_{ij}/\lambda)],
 \end{equation}
and the corresponding modified gravitational forces are
 \begin{equation}
	 \textrm{F}_i = -\sum^N_{j \ne i}\frac{m_i m_j G}{r_{ij}^2}[1+\alpha exp(-r_{ij}/\lambda)(1+r_{ij}/\lambda)] 
	  \sim  -\sum^N_{j \ne i}\frac{m_i m_j G}{r_{ij}^2}. 
 \end{equation}
 The Yukawa  correction Eq. (16) modifies the attraction at medium distances, and goes asymptotically over to the classical gravitational
forces at large distances. The range of modification is determined by the parameters $\lambda$ and modification strength $\alpha$ with
 \begin{equation}
    \lambda/\alpha=  r_0=\sqrt{m_jG/a_0}
 \end{equation}

 \begin{figure}[!]
 \begin{center}	  
 	 \includegraphics[width=6cm,angle=-90]{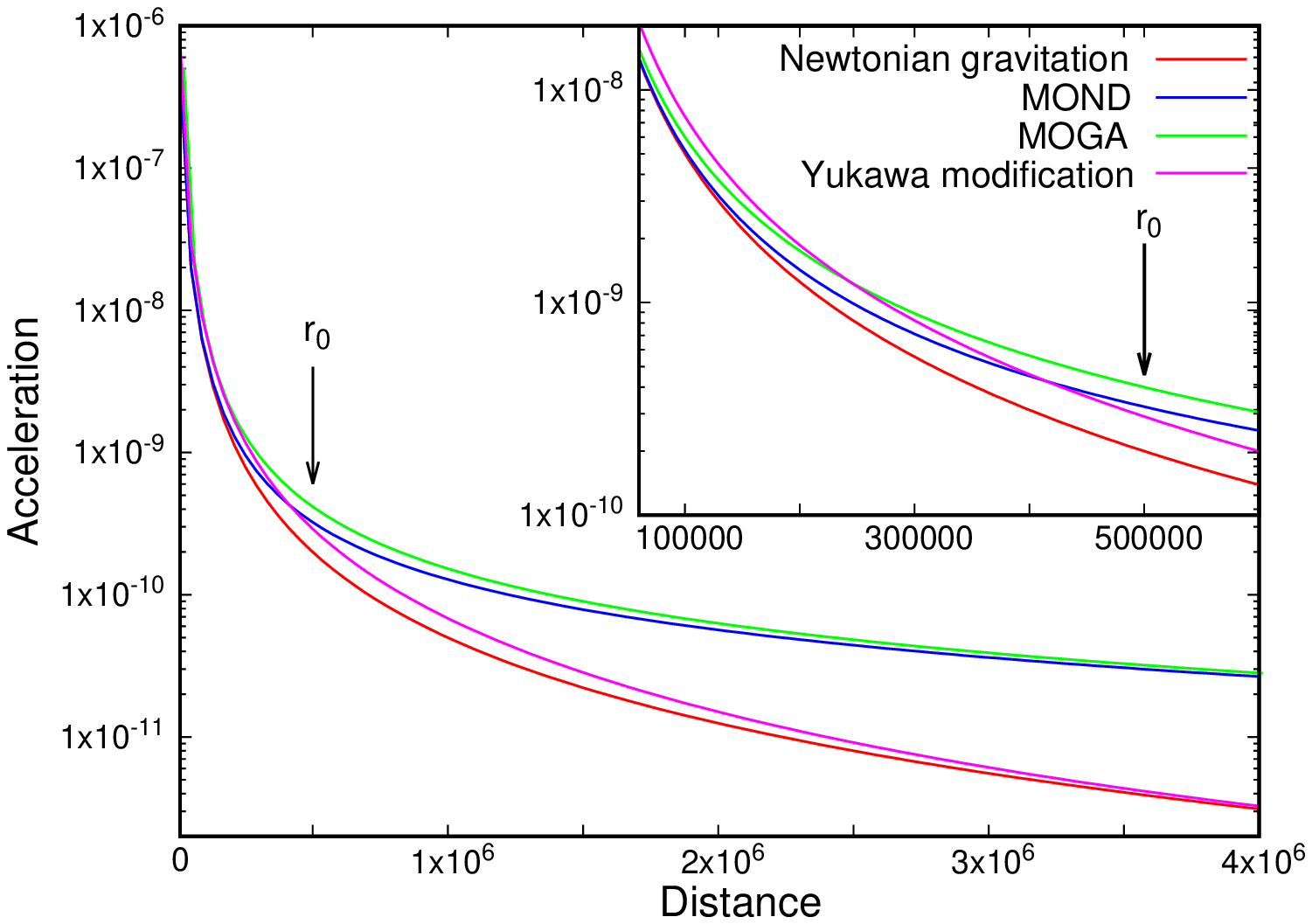}
	  \caption{Accelerations of the outer star  (No. 2 with blue in Figure 1) as a
	  function of the distance to object No. 3, and for the  
	  modifications of the acceleration (MOND) and the gravitational attraction (MOGA and Yukawa).
	  Red curve is the classical Newtonian acceleration. The MOND modification of the acceleration is in blue. The MOGA  
	  modification is in green,
	  and the Yukawa modification is in magenta.}
 \end{center}	
\end{figure}

The modified gravitational attraction, Eq. (16), is  an example of a general modification of the force field. 
 The parameters $\lambda/ \alpha$ corresponds to the parameter $r_0$ in MOGA, and 
   a possible deviation from the ISL gravity was investigated for $\lambda$ in
   the range $\lambda \in [30,8000]$ nm  \cite{Chen2016,Bimonte2021,Baeza-Ballesteros2022}, and the rotation velocities of stars in the Milky Way
   were used to determine a possible  Yukawa correction, Eq. (16) to the gravitational attraction \cite{Henrichs2021}. So far, however, there is no
   direct experimental evidence for deviation from the ISL gravitational attraction. 

\subsection{Testing  MOND-, Yukawa-, and MOGA-modifications on the TBS}

The stability of the regular dynamics in TBS with modified accelerations or gravitational attractions has been studied for the
MOND, Yukawa, and MOGA modifications.
All three modifications increase the accelerations of objects at large distances.
The different modified accelerations for the interaction with one object are shown in Figure 4.
 The accelerations are for the object No. 2 with mass $m_2=0.5 $ attracted by the heavy mass center No. 3 with
a mass  $m_3=100 $   that is two hundred times heavier than the object. The acceleration is
equal to the acceleration the object No. 2 obtains from the attraction of object No. 3 in the TBS.

The classical Newtonian ISL acceleration is in red in Figure 4, and the corresponding accelerations
for the various modifications with  $r_0=500000$ are shown in blue for MOND, 
with magenta for the Yukawa modification Eq. (16), and in green for MOGA. The MOND and MOGA are asymptotically equal, and the classical ISL and Yukawa potential
agree correspondingly asymptotically for large distances.
The inset in the figure enlarges the accelerations at distances $r \le r_0$.

Models of  galaxies with  classical Newtonian dynamics have a Keplerian rotation velocity of the stars, which declines
with the square root of the distance to the center of rotation. This is in disagreement with experimentally determined 
velocities of galaxies, which are rather constant with respect to the distance to the center of rotation \cite{Gentile2011}.
This motivated Milgrom  to formulate the MOND modification with the increased acceleration at large distances 
to overcome this shortcoming, and the enhanced accelerations of all three modifications result in a modified rotation velocity for galaxial
systems in qualitative agreement with rotations of spiral galaxies in the Universe,
MOND: \cite{Gentile2011,Pflamm-Altenburg2025a}; Yukawa: \cite{Brandau2012}; and MOGA:\cite{Toxvaerd2024a}.

 The  impact of PM approximation, shown in Figure 3a and Figure 3b,  is for distances
 $r > r_0=500000$, and  the   results reported below for the dynamics with MOND, Yukawa, and MOGA modification are correspondingly for $r_0=500000$, i.e.
  MOND: $a_0= r_0^{-2}= 4 \times 10^{-10}$, and Yukawa: $\lambda=$ 500000.

 \begin{figure}\label{sec4}
	   \begin{center}
 	 \includegraphics[width=6cm,angle=-90]{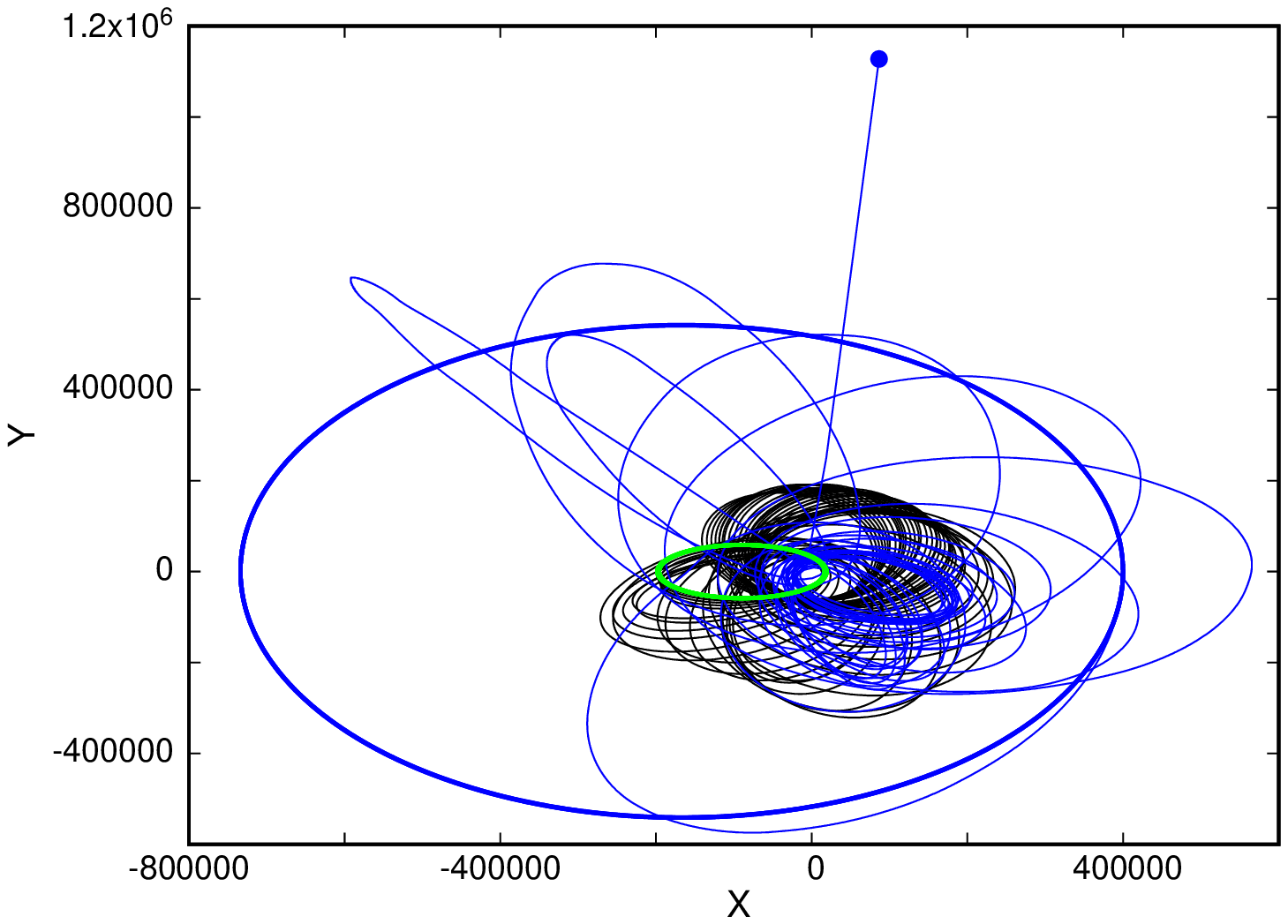}
	   \caption{The MOND modification for $a_0=4 \times 10^{-10 }$ (corresponding to $r_0=500000$ in Yukawa,MOGA, and PM).
	   The orbits with MOND dynamics are shown in thin blue for No. 2, and in thin black for No. 1.
	   The position of No. 2 at  $t=1.2576 \times 10^9$  at its release is marked with a blue sphere.
	   The ellipses with thick blue and green (from Fig. 1) are without MOND  modification.}
	   \end{center}
 \end{figure}

 \subsubsection{MOND modification of the TBS}
 The acceleration of one object with MOND, shown with blue in Figure 4,
is increased already for distances significantly less than $r_0=500 000$, and the acceleration
for the improved QuMOND increases the acceleration further
\cite{Pflamm-Altenburg2025a,Pflamm-Altenburg2025b,DiCintio2025}. However,
the increased accelerations affect the regular dynamics of the objects in TBS and their long-time stability. 
Figure 5 shows the impact of MOND on the regular dynamics in the TBS.
  The orbits are for $1.2576 \times 10^7$ time steps with $\delta t=100$, and
  at the time where object No. 2 (in blue) is released from the TBS. But  also the dynamics of No. 1
  nearby the center of gravity is affected by the
 MOND modification of the acceleration. Its orbits are shown in black in the figure.

 Simulations with other values of $a_0$ and other TBS gave the same qualitative result.
The regular dynamics in TBS is sensitive to the MOND modification of the acceleration,
with a destruction of the regular dynamics and the release of the objects.
The destruction of the regular dynamics is due to the violation of Newton's third law (Eq. 11),
which demolishes the momentum conservation \cite{Felten1984,Toxvaerd2024a}.

\begin{figure}
\begin{center}	  
\begin{subfigure}[a]{1.00\linewidth}
\begin{center}	 
 	 \includegraphics[width=6.cm,angle=-90]{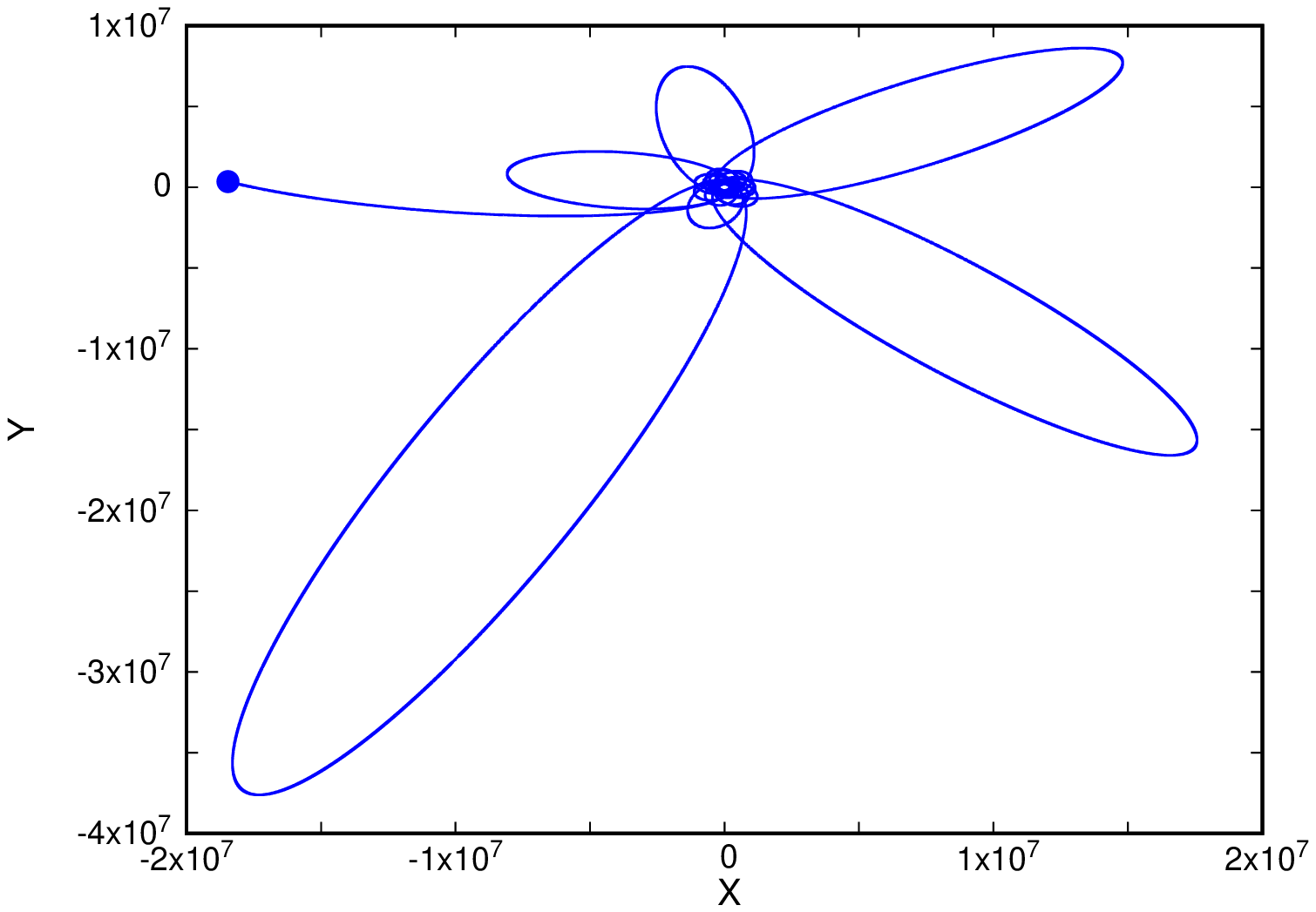}
	 \caption{ The orbit of No. 2 with the Yukawa modification with $\lambda$=500000, and $\alpha=1$. The object (blue sphere) is accelerated out of the TBS
	 at time $t=10^{11.1}$.}
 \end{center}	
\end{subfigure}

\begin{subfigure}[b]{1.00\linewidth}	  	  
\begin{center}	
 	 \includegraphics[width=6.cm,angle=-90]{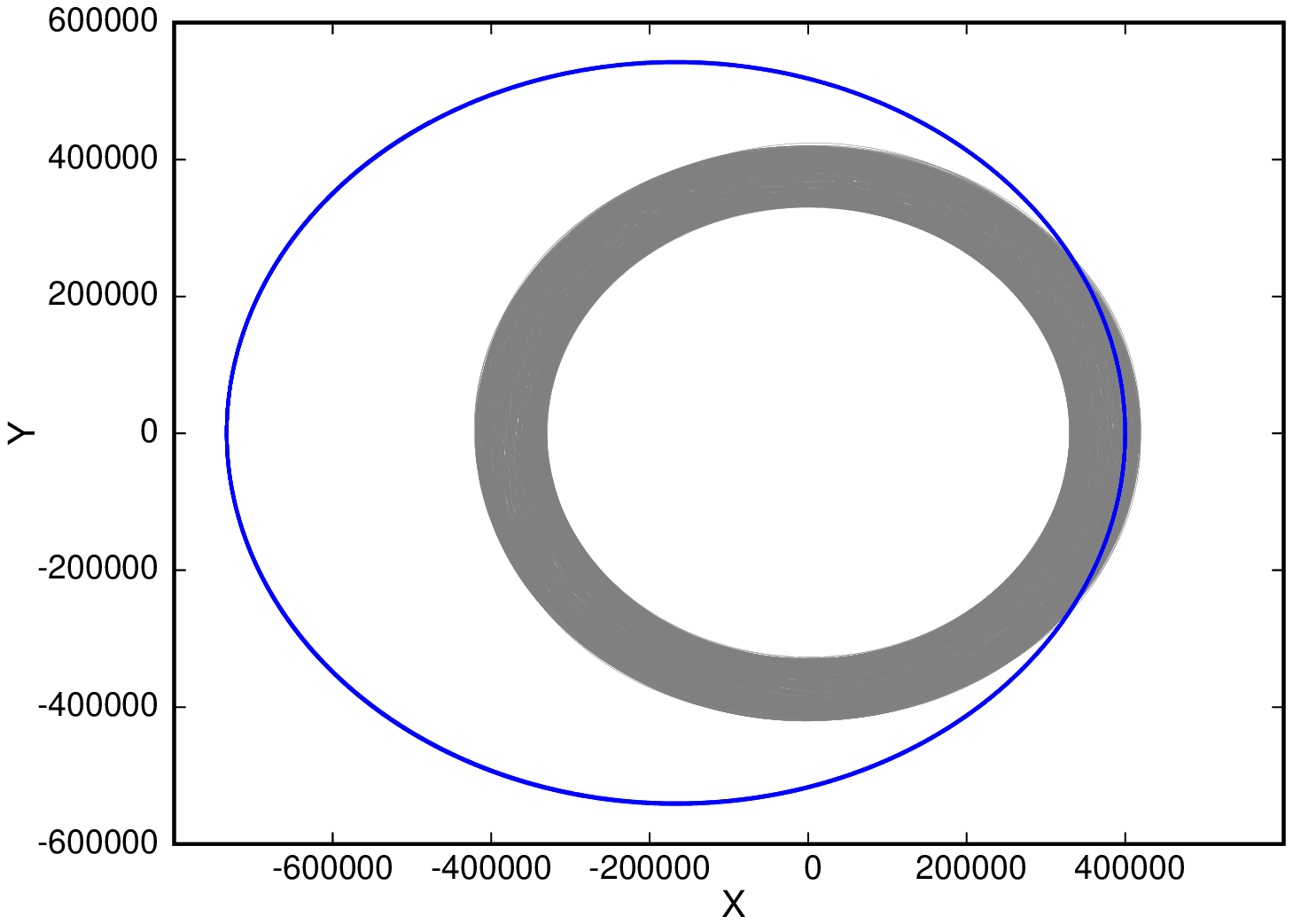}
	 \caption{ The  corresponding $\approx$ 800 stable regular revolving orbits of No. 2  with light grey for
	 $\lambda$=500000, and $\alpha=0.5$. The orbit in blue is the regular orbit from Figure 1 without the Yukawa modification.}
\end{center}	
\end{subfigure}	 
	  \end{center}
	  \caption{ Simulation with the Yukawa modification  for $\lambda=500000$.}
\end{figure}

 \subsubsection{Yukawa mdification of the TBS}
The Yukawa modification is a modification of the force from a far-away object in the galaxy,
and a modification of the force between pairs of objects obeys Newton's third law and maintains all the invariances of classical mechanics,
and the regular dynamics in the TBS system is, in general, preserved.

 The Yukawa potential Eq. 16 contains two parameters $\alpha$ and $\lambda$, where  $\alpha$ is the intensity of the exponential
 declining modification, and  $\lambda$ is a measure for the range of the modification, and the
 net modification is given by the
  ratio $\lambda/\alpha= r_0$. Figure 6a and 6b shows the regular dynamics for two values of the intensity $\alpha$
  of the Yukawa modification.  The magenta curve in Figure 4 is for $\alpha=1$, and $\lambda=500 000$, and it shows the increased acceleration on the
  objects from the heavy center $M=m_3=100$ for $r \le r_0$.
  But the increased accelerations can accelerate an object out of the TBS model of a dwarf galaxy, which happened for $\alpha=1$ and
  with $\lambda=500000$.  The event is shown in Figure 6a, where object No. 2 is accelerated out of the TBS system after
  $10^{9.1}$ timesteps. However,  the Yukawa accelerations result in stable regular orbits,  for a reduced
  intensity of the Yukawa modification, which are shown in 
  Figure 6b for $\alpha=0.5$ and $\lambda=500000$. 

The TBS system was simulated for other values of $\alpha,\lambda$, and the Yukawa modification will, in general, stabilize the regular dynamics.

\begin{figure}
\begin{center}	  
\begin{subfigure}[a]{1.00\linewidth}
\begin{center}	 
 	 \includegraphics[width=6.cm,angle=-90]{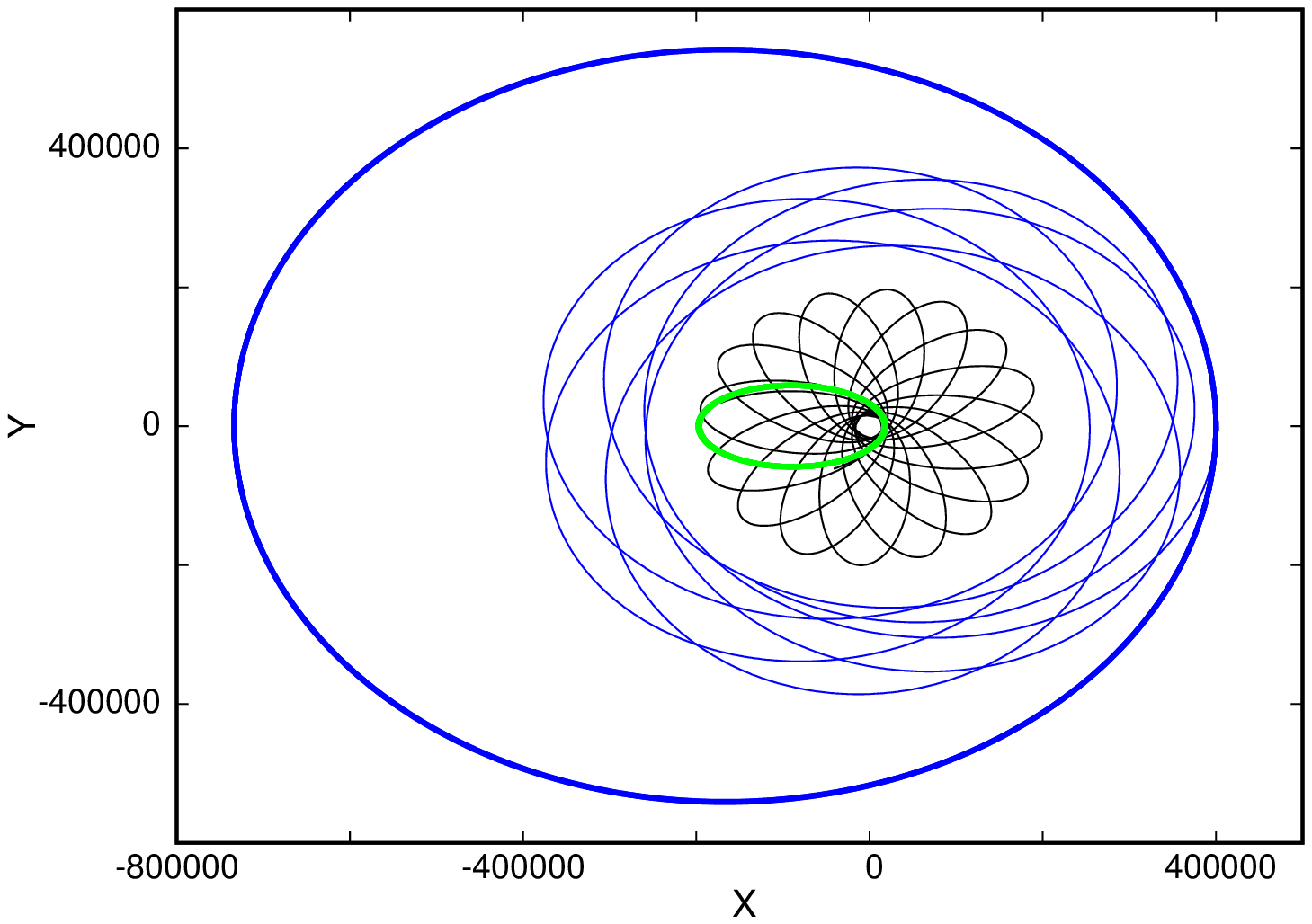}
	 \caption{ The start of MOGA. The revolving elliptical orbits are shown in black for No.1  and blue for No. 2, together with the elliptical
	 orbits without MOGA ( from Figure 1) with thick lines. The simulations are for the times where the elliptical orbits have revolved 2$\pi$ .}
 \end{center}	
\end{subfigure}

\begin{subfigure}[b]{1.00\linewidth}	  	  
\begin{center}	
 	 \includegraphics[width=6.cm,angle=-90]{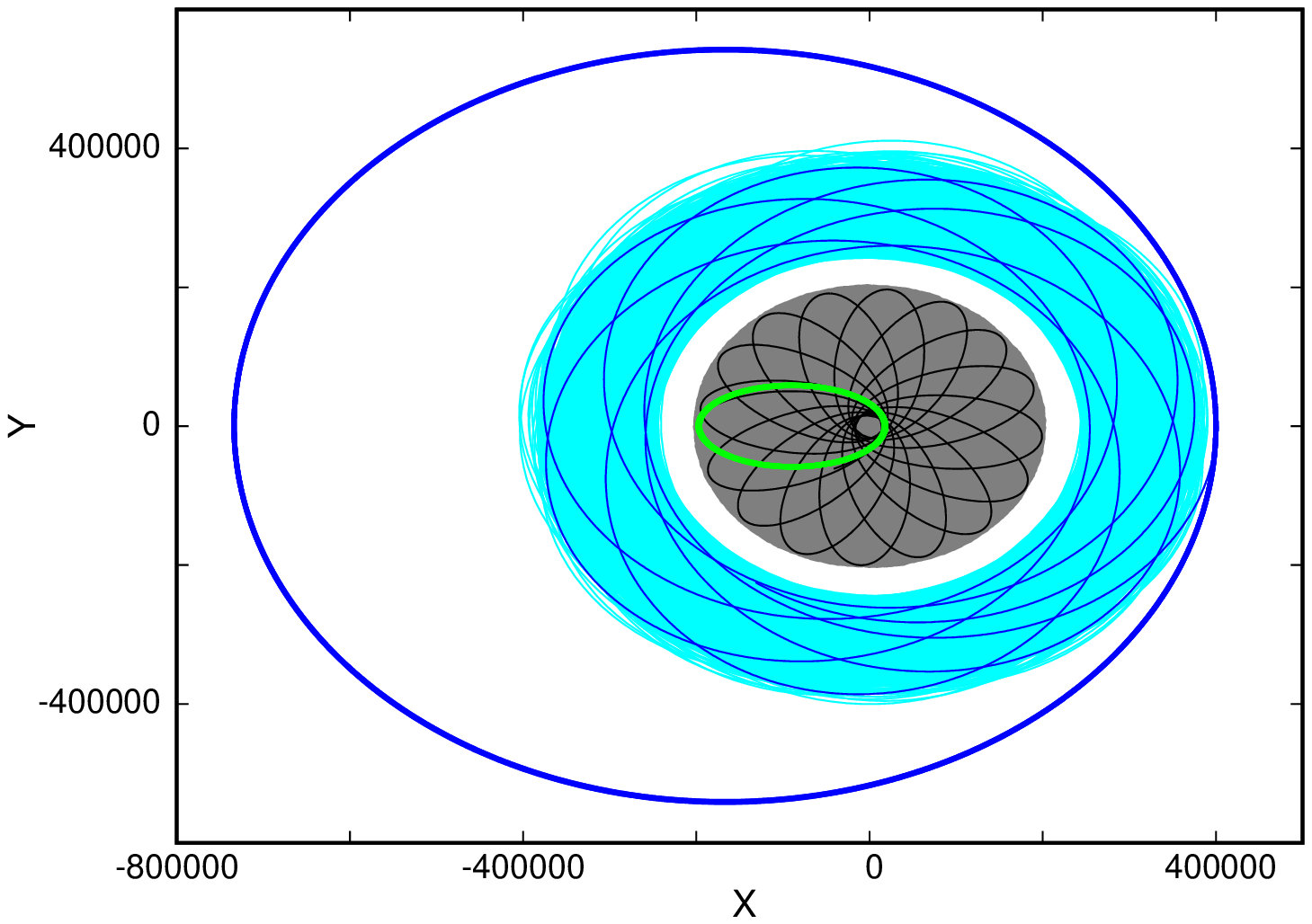}
	 \caption{ The revolving orbits with MOGA and for $10^9$ timesteps. The revolving elliptical orbits with gray for No. 1 and in light blue for No. 2 
	 are shown together with the orbits from Figure 6(a). }
\end{center}	
\end{subfigure}	 
	  \end{center}
	  \caption{ Simulation with MOGA and for $r_0=500000$.}
\end{figure}

 \subsubsection{MOGA modification of the TBS}
 The corresponding impact of the MOGA modification of the attractions is shown in Figure 7. The dynamics are for the modification distance $r_0=500000$,
 as in PM and the previous modifications. Similar to the Yukawa modification, the modification of the gravitational attractions from an inverse
 square attraction to an inverse attraction conserves the classical dynamics invariances, and MOGA stabilizes the regular dynamics,
 but with ``revolving'' elliptical orbits \cite{Toxvaerd2022}. Figure 7a shows the orbits
 after the revolving orbits have turned  $2 \pi$ around. For  No.1 (in black) after $t_1= 3.155 \times 10^6$ timesteps,
 and after $t_2 = 2 \times t_1= 6.310 \times 10^6$ time steps for No. 2 (in thin blue). The revolving orbits are with conservation of the length of the principal axis. 
 Figure 7b shows the revolving orbits after $10^9$ timesteps, with the orbits of No. 1 in gray and light blue for No. 2. 

Similar to the Yukawa modification, TBS was simulated with other value of $r_0$ and for other TBS systems,
and with the general result that MOGA stabilizes the regular dynamics.\\

$ $\\

$ $\\

\newpage

\section{Summary of the tests, and the conclusion}
The purpose of this article is to investigate the effect of the  PM approximations and the modifications of forces and accelerations,
used in simulations of galaxies, on their regular dynamics.
The three-body system (TBS) is the simplest celestial system to test the approximations and modifications used in celestial dynamics,
and it is easy to implement on a computer (see Appendix).
\subsection{Summary of the tests}
 Galaxies simulated with classical Newtonian dynamics and with PM  or MOND are unstable, and the exact simulations of the TBS system show that
the PM approximation, as well as the MOND modification of the accelerations, destabilize the regular dynamics.
In contrast, the Yukawa and MOGA modifications with an increased 
gravitational attraction stabilize the TBS and its regular dynamics. The break of momentum conservation
 by the PM approximations and the MOND modification, caused by a violation of Newton's third law, most likely causes the destabilization of the regular dynamics.
However, all the modifications, MOND, QuMOND, Yukawa, and MOGA, increase the acceleration
of far-away objects and all the modifications exhibit velocity profiles in qualitative agreement with the experimentally determined rotation velocities
of galaxies.
\subsection{Conclusion}
 Classical celestial dynamics for an $N$-body system have regular stable orbits.
 However, an extrapolation from a three-body system to a galaxy with a hundred billion stars
must be taken cautiously, and only additional simulations can reveal
 the stability of galaxies without approximations or with modifications. But large-scale simulations without approximation are extremely time-demanding.

All of the modifications, MOND, QuMOND, Yukawa, and MOGA, increase the acceleration
of distant objects, and all of the modifications exhibit velocity profiles in qualitative agreement with the experimentally determined rotation velocities
of galaxies MOND: \cite{Gentile2011,Pflamm-Altenburg2025a}; Yukawa: \cite{Brandau2012}; and MOGA: \cite{Toxvaerd2024a}.
The modifications are $\textit{ad hoc}$, and only improved analyses and theories can determine whether an asymptotic modification of gravity
is consistent with the current foundation of physics.
In \cite{Toxvaerd2024a}, we proposed that the modification of the ISL attractions is caused by a lensing and focusing by the
gravitational objects in the central part of the galaxy of the gravitational waves
from distant objects. A lensing that will act as a self-stabilizing effect on the halos of galaxies.

\section{Acknowledgment}
This work was supported by the VILLUM Foundation’s Matter project, grant No. 16515.\\
$ $\\
$\textbf{Data availability}$---  Data and computer programs will be available on request.

\section{ Appendix }

\subsection{Newton's Discrete algorithm}
Almost all simulations of celestial systems are with Newton's algorithm for discrete dynamics.
  In Newton's  discrete dynamics \cite{Newton1687} the time and forces are discrete with discrete force
  impulses at every discrete times $t, t+\delta t, t+2\delta t,...$,.  A new  position $\textbf{r}_i(t+\delta t)$ at time $t+\delta t$ of an object
$i$ with the mass $m_i$  is determined by
the force impulse  $\textbf{f}_i(t)$ at time $t$ acting on the object   at the  positions $\textbf{r}_i(t)$, and  
 the position $\textbf{r}_i(t-\delta t)$ at $t - \delta t$  as
\begin{equation}
	 m_i\frac{\textbf{r}_i(t+\delta t)-\textbf{r}_i(t)}{\delta t}
			=m_i\frac{\textbf{r}_i(t)-\textbf{r}_i(t-\delta t)}{\delta t} +\delta t \textbf{f}_i(t),	
 \end{equation}
where the momenta $ \textbf{p}_i(t+\delta t/2) =  m_i (\textbf{r}_i(t+\delta t)-\textbf{r}_i(t))/\delta t$ and
 $  \textbf{p}_i(t-\delta t/2)=  m_i(\textbf{r}_i(t)-\textbf{r}_i(t-\delta t))/\delta t$ are constant in
the time intervals in between the discrete positions.

 Usually, the algorithm, Eq. (1), is  presented  as the Leap-frog algorithm for the velocities
\begin{equation}
\textbf{v}_i(t+\delta t/2)=  \textbf{v}_i(t-\delta t/2)+ \delta t/m_i  \textbf{f}_i(t),
\end{equation}
 and the positions
are determined from the discrete values of the momenta/velocities as
\begin{equation}
\textbf{r}_i(t+\delta t)= \textbf{r}_i(t)+ \delta t \textbf{v}_i(t+\delta t/2).	  
\end{equation}

The discrete algorithm is time-reversible due to the time symmetry,
which also ensures the symplecticity and the energy conservation \cite{Toxvaerd2023},
and Newton's third law
\begin{equation}
	\textbf{f}_{ij}(t)=-\textbf{f}_{ji}(t)
\end{equation}	
ensures the  conservation of momentum and angular momentum.

\subsection{The three-body system}
 The TBS consists of three objects with masses  $m_1, m_2$, and $m_3$.
 The Newtonian discrete dynamics with time increment $\delta t$  is obtained
 from three positions $\textbf{r}_1(t),  \textbf{r}_2(t),$ and $\textbf{r}_3(t)$ 
and  three veloceties  $\textbf{v}_1(t-\delta t/2),  \textbf{v}_2(t-\delta t/2),$ and
$\textbf{v}_3(t-\delta t/2)$, in total six three-dimensional dynamic variables. The present TBS is started at time $t=0$ with
all three objects in their aphelion or perihelion and with start velocities in the  ``Ècliptica'' plane,
given by the plane with the three objects' positions at the start.  The three objects remain in the plane
since the discrete dynamics conserves momentum, and the system is two-dimensional (2D), and
with three (x,y)-positions and their velocities, in total, twelve dynamic values which need to be specified at
the start of the simulation.
The conserved momentum and center of mass
reduces the number of necessary start values to eight.

The TBS exhibits a variety of regular dynamics depending on the start values, for a review see  \cite{Krishnaswami2019}.
The present TPS system is created with two objects in elliptical orbits around a third heavy object.
The objects' 
six  components of position $x_1(0)$, $ y_1(0),$ $x_2(0),$ $y_2(0),$ $x_3(0),$$ y_3(0)$ and
six velocety components  $vx_1(-\delta t/2)),$ $vy_1(-\delta t/2)),$ $vx_2(-\delta t/2)),$
$vy_2(-\delta t/2)),$ $vx_3(-\delta t/2)),$ $vy_3(-\delta t/2))$  must be specified.

Aphelion or perihelion, with the longest of the principal axes in the x-direction, determines
 six  start values:

\begin{equation} 
	y_1(0)= y_2(0)= y_3(0)=0,
\end{equation}
and with
\begin{equation} 
 vx_1(0)= vx_2(0)=vx_3(0)=0.
\end{equation}
We need to know the velocities, not at time  $t=0$, but at time $-\delta t/2$.
However, these velocities can be calculated by the algorithm and the force  in the x-direction $ \textbf{f}_{x}(0)$  using time symmetry since
\begin{equation}
	vx_1(\delta t/2)=-vx_1(-\delta t/2) ( \textrm{time reversibility)},
\end{equation}	
and (Eq. 16)
\begin{equation}
	vx_1(\delta t/2)=vx_1(-\delta t/2)+ \delta t/m_1 \textbf{f}_{x_1}(0),
\end{equation}
by which
\begin{equation}
	vx_1(-\delta t/2)= -\delta t \textbf{f}_{x_1}(0)/2m_1 ,
\end{equation}
and correspondingly for the two other objects. 

The conserved momentum and center of mass give two relations.
Let the conserved center of mass be the origin of the 2D coordinate system, then
\begin{equation}
	m_1x_1(t)+m_2x_2(t)+m_3x_3(t)=0,
\end{equation}
and the conserved momentum gives
\begin{equation}
	m_1vy_1(t)+m_2vy_2(t)+m_3vy_3(t)=0.
\end{equation}
So the dynamics of the TBS system with the tree objects in their aphelion or perihelion are obtained by specifying four values.

 The TBS system with the ellipses in Figure 1, used in the present investigations,
 is started with (mass unit given by $m_1$ and dynamics units given by the
 gravitational constant $G=1$) with \\
$ input: m_1, x_1(0), vy_1(-\delta t/2), m_2, m_3$ = 1, -200000, 0.009, 0.5, 100,\\
and with the heavy object at the origin and in rest at the start
 $vx_3(-\delta t/2)=vy_3(-\delta t/2)=0$. The figures in the article are obtained by the discrete algorithm with $\delta  t$=100. The result with regular dynamics, shown in
 the figures is not sensitive to  $\delta  t$.

\end{document}